%% file: 19SPAWC-1bTimeFreqSync_arxiv.tex
\documentclass[10pt,conference,twocolumn]{IEEEtran}

\usepackage{cite}
\usepackage[T1]{fontenc}
\usepackage{graphicx}
\usepackage{amssymb}
\usepackage{amsmath}
\usepackage{paralist}
\usepackage{setspace}
\usepackage{microtype}
\usepackage{balance}
\usepackage[caption=false,font=footnotesize]{subfig}
\usepackage{xcolor}
\usepackage{accents}
\usepackage{dsfont}
\usepackage{lipsum}
\usepackage{booktabs}
\usepackage[nolist,nohyperlinks]{acronym}

\newcommand{\quantize}{\mathcal{Q}}

\newcommand{\snr}{\rho}

\input{vmr-symbols-vecbold}
\input{standard-macros}

\input{defs}

\IEEEoverridecommandlockouts
\allowdisplaybreaks
\begin{document}

%
\title{Timing and Frequency Synchronization for \\ 1-bit Massive MU-MIMO-OFDM Downlink}
\author{\IEEEauthorblockN{Sven Jacobsson$^\text{1,2}$, Carl Lindquist$^\text{1,2}$, Giuseppe Durisi$^\text{1}$, Thomas Eriksson$^\text{1}$, and Christoph Studer$^\text{3}$} \\[-0.3cm]
\thanks{The work of SJ and GD was supported in part by the Swedish Foundation for Strategic Research under grant ID14-0022, and by the Swedish Governmental Agency for Innovation Systems (VINNOVA) within the competence center ChaseOn. The work of CS was supported in part by Xilinx, Inc.~and by the US National Science Foundation under grants ECCS-1408006, CCF-1535897,  CCF-1652065, CNS-1717559, and ECCS-1824379.}
\IEEEauthorblockA{
\small $^\text{1}$\textit{Chalmers University of Technology, Gothenburg, Sweden};  
\,$^\text{2}$\textit{Ericsson Research, Gothenburg, Sweden}; 
\,$^\text{3}$\textit{Cornell University, Ithaca, NY}
}
}

\maketitle

\begin{abstract}

We consider timing and frequency synchronization for the massive multiuser (MU) multiple-input multiple-output (MIMO) downlink where 1-bit digital-to-analog converters~(DACs) are used at the base station (BS).
We focus on the practically relevant scenario in which orthogonal-frequency division multiplexing (OFDM) is used to communicate over frequency-selective channels.
Our contributions are twofold. 
First, we use Bussgang's theorem to analyze the impact on performance caused by timing and frequency offsets in the presence of 1-bit DACs at the~BS.
Second, we demonstrate the efficacy of the widely used Schmidl-Cox synchronization algorithm. 
Our results demonstrate that the 1-bit massive MU-MIMO-OFDM downlink is resilient against timing and frequency~offsets. 
%

\end{abstract}

\section{Introduction} \label{sec:intro}


Massive multiuser~(MU) multiple-input multiple-output (MIMO) will be a key technology component in fifth-generation (5G) radio access networks~\cite{larsson14a}.
Among its advantages, massive MU-MIMO overcomes the strong path  loss at millimeter-wave (mmWave) frequencies by means of beamforming~\cite{swindlehurst14a}. 
In order to fully exploit the large bandwidth available at mmWave frequencies with all-digital beamforming architectures  that use homodyne radio transceivers, a pair of high-speed digital-to-analog converters (DACs) is required at each antenna port.
Such all-digital architectures for massive MU-MIMO mmWave systems require innovative hardware solutions to limit power consumption, interconnect bandwidth, and system costs. One of the most promising approaches is to lower the resolution of the DACs~\cite{boccardi14a}.

\subsection{Previous Results}

The use 1-bit DACs in the massive MU-MIMO downlink has been investigated in, e.g.,~\cite{saxena16a, jacobsson17d, castaneda17a, li17a, sohrabi18a} for transmission over frequency-flat  channels, and in, e.g.,~\cite{jacobsson17e, nedelcu17a, jacobsson17f} for transmission over frequency-selective channels with orthogonal frequency-division multiplexing (OFDM). 
All of these results show that high throughput at low bit-error rate (BER)  can be achieved despite the quantization artifacts introduced by the 1-bit DACs.
It is, however, an open question whether sufficiently accurate timing and frequency synchronization can be achieved in such low-resolution BS architectures.


%
For the infinite-resolution (quantization-free) case, timing and frequency synchronization for OFDM systems, which involves retrieving the symbol-timing offset (STO) and the carrier-frequency offset~(CFO), has been studied extensively in the literature; see, e.g.,~\cite{morelli07a} for a review.

Under the assumption of perfect frequency synchronization, algorithms for estimating the STO in the presence of 1-bit measurements have been presented in~\cite{beek95b, stein17a, schluter18a}. 
Furthermore, under the assumption of perfect timing synchronization, algorithms for estimating the CFO using coarsely quantized measurements have been proposed in~\cite{wadhwa13a, myers17a}. 
For wideband mmWave systems that use low-resolution analog-to-digital converters, timing and frequency synchronization algorithms based on Zadoff-Chu sequences have been discussed recently in~\cite{zhu18a, zhu18b}. 
All these results, however, are limited to the case of 1-bit quantization at the receiver. In this work, we shall consider timing and frequency synchronization in the presence of 1-bit quantization at the transmitter.




\subsection{Contributions and Outline}
In contrast to existing results, we investigate joint timing and frequency synchronization 
 in the massive MU-MIMO-OFDM downlink where 1-bit DACs are used at the BS. 
In~\fref{sec:system_model}, we introduce the system model.
In~\fref{sec:analysis}, we use Bussgang's theorem~\cite{bussgang52a} to analyze the impact on performance of residual timing and frequency offsets.
In~\fref{sec:synch}, we demonstrate that the Schmidl-Cox algorithm~\cite{schmidl97a} successfully achieves timing and frequency synchronization despite the use of 1-bit DACs at the BS.
We conclude the paper in~\fref{sec:conclusions}.

%
%


\section{System Model} \label{sec:system_model}

We consider a massive MU-MIMO-OFDM downlink system in which $B$ BS antennas serve $U$ single-antenna user equipments (UEs).  
Throughout the paper, we assume that each BS antenna element is fed by a pair of 1-bit DACs. 
Our model includes timing and frequency offsets between the BS and the UEs.
The timing offset is caused by an unknown frame start instant and by propagation delays; the frequency offset is caused by oscillator instabilities and Doppler~shift.

\subsection{Channel Input-Output Relation} \label{sec:channel_in_out}

In the presence of timing and frequency offsets, the $n$th sample ($n \in \opZ$) of the time-domain signal received at the $u$th UE can be written as
\begin{IEEEeqnarray}{rCl} \label{eq:y_freq_selective}
	y_u[n] &=& e^{-\frac{j2\pi{\varepsilon_u}n}{N}} \sum_{\ell = 0}^{L-1}\vech_u^T[\ell] \quantize\lefto(\vecx[n-\ell-\tau_u]\right) + w_u[n] \IEEEeqnarraynumspace
\end{IEEEeqnarray}
for $u = 1, 2, \dots, U$. 
Here, $\varepsilon_u \in \opR$ is the CFO at the $u$th UE (normalized by the subcarrier spacing), $\tau_u \in \opZ$ is the STO at the $u$th UE (which we model as integer-valued, since fractional STO can be absorbed into the impulse response of the channel), $w_u[n] \distas \jpg(0,N_0)$ is the UE-side AWGN, and $\vech_u[\ell] \in \opC^B$ is the $\ell$th tap ($\ell = 0,1,\dots,L-1$) of the $B$-dimensional channel between the BS and the $u$th UE.
In this work, we consider Rayleigh fading with a uniform power delay profile. 
Specifically, the elements of $\{\vech_u\}$ are independently drawn from a $\jpg(0, 1/L)$ distribution. 
The nonlinear function $\quantize(\cdot): \opC^B \rightarrow \setX^B$, where $\setX = \sqrt{{1}/{(2B)}}\{ 1+j, -1+j, -1-j, 1-j\}$, describes the joint operation of the $2B$~1-bit DACs at the~BS:
\begin{IEEEeqnarray}{rCl} \label{eq:quantize}
	\quantize\lefto(\vecx[n]\right) &=& \sqrt{\frac{1}{2B}} \Big(\!\sign\lefto( \Re\{\vecx[n]\}\right) + j\sign\lefto( \Im\{\vecx[n]\}\right)\!\Big). \IEEEeqnarraynumspace
\end{IEEEeqnarray}
The 1-bit DACs ensure that $\opnorm{\quantize(\vecx[n])}^2 = 1$ for every $\vecx[n]$. 

\subsection{OFDM Processing and Linear Precoding}

The time-domain precoded vector $\vecx[n] \in \opC^B$ in~\eqref{eq:y_freq_selective} is given by $\vecx[n] = \sum_{i}\vecx^{(i)}[n-i(N+G)]$, where $\vecx^{(i)}[n]$ is the time-domain precoded vector for the $i$th ($i \in \opZ$) OFDM symbol, which is obtained by computing $B$ size-$N$  inverse DFTs:
\begin{IEEEeqnarray}{rCl} \label{eq:x_m_time}
\vecx^{(i)}[n] &=& 
\begin{cases}
\frac{1}{\sqrt{N}} \sum\limits_{k=0}^{N-1} \hat\vecx^{(i)}[k] e^{\frac{j2\pi{k}n}{N}}, & \!\!\!-G \le n \le N-1\\
0, & \!\!\text{otherwise}.
\end{cases} \IEEEeqnarraynumspace
\end{IEEEeqnarray}
Here, $\hat\vecx^{(i)}[k]$ is the corresponding frequency-domain precoded vector and $G \ge L-1$ is the length of the cyclic prefix (CP).
We assume that linear precoding is used at the BS, which implies that $\hat\vecx^{(i)}[k]$ can be written~as
\begin{IEEEeqnarray}{rCl} \label{eq:x_m_freq}
	\hat\vecx^{(i)}[k] = \sum_{u=1}^U\hat\vecp_u[k]\hat{s}_u^{(i)}[k]
\end{IEEEeqnarray}
for $k = 0,1,\dots,N-1$.
Here $\hat{s}_u^{(i)}[k] \in \opC$ is the frequency-domain symbol at the $k$th subcarrier intended for the $u$th UE during the $i$th OFDM symbol.
We use $s_u^{(i)}[n] = \frac{1}{\sqrt{N}}\sum_{k=0}^{N-1} \hat{s}_u^{(i)}[k] e^{\frac{j2\pi{k}n}{N}}$ for $n = -G, -G+1, \dots, N-1$ to~denote the corresponding time-domain symbols.
Furthermore, $\hat\vecp_u[k] \in \opC^B$ is the frequency-domain precoding vector used to map the symbols on the $k$th subcarrier intended for the $u$th UE to the BS antenna~array.

Typically, not all subcarriers are used for symbol transmission. Let~$\setS$ denote set of used subcarriers and let $S = \abs{\setS} \le N$ denote the number of used subcarriers. 
We assume that $\Ex{}{|\hat{s}_u[k]|^2} = 1$ for $k \in \setS$ and $\hat{s}_u[k] = 0$ for $k \notin \setS$. 
%
%
We define the oversampling ratio~(OSR) as $\textit{OSR} = N/S$.
%

Timing and frequency synchronization involves estimation and compensation of STO and CFO.
At the $u$th UE, the STO and CFO are compensated for in the time domain using the estimates $\tau_u^\text{est}$ and $\varepsilon_u^\text{est}$ of $\tau_u$ and $\varepsilon_u$, respectively. 
In~\fref{sec:synch}, we will discuss how to obtain such estimates using the well-known Schmidl-Cox algorithm.
The time-domain received signal after timing and frequency synchronization is
\begin{IEEEeqnarray}{rCl} \label{eq:compensated_time}
	r_u[n] = e^{\frac{j2\pi\varepsilon_u^\text{est}n}{N}} y_u[n + \tau_u^\text{est}].
\end{IEEEeqnarray}
The corresponding frequency-domain signal received on the $k$th subcarrier ($k \in \setS$) during the $i$th DFT window is given~by
\begin{IEEEeqnarray}{rCl} \label{eq:compensated_freq}
\hat{r}_u^{(i)}[k] &=& \frac{e^{\frac{j\pi G k}{N}}}{\sqrt{N}} \sum_{n=0}^{N-1}	r_u^{(i)}[n]  e^{-\frac{j2\pi k n }{N}} \IEEEeqnarraynumspace
\end{IEEEeqnarray}
where
\begin{IEEEeqnarray}{rCl} \label{eq:tilt}
	r_u^{(i)}[n] = r_u[n + i(N+G) - G/2].
\end{IEEEeqnarray}
Let $\Delta\tau_u = \tau_u^\text{est} -\tau_u$ and $\Delta\varepsilon_u = \varepsilon_u^\text{est} -\varepsilon_u$ denote the \emph{residual} STO and CFO, respectively.
Clearly, if timing synchronization is not perfect and there is a residual STO between the BS and the $u$th UE ($\Delta\tau_u \neq 0$), the DFT window will be placed in an incorrect position, which may cause inter-carrier interference (ICI) and inter-symbol interference~(ISI).
Residual CFO due to imperfect frequency synchronization may introduce further~ICI due to the loss of orthogonality between subcarriers.
In~\eqref{eq:tilt}, the DFT window is shifted to the left by $G/2$ samples. By doing so, ISI-free communication is achieved if the residual STO is $-G/2+L-1 \le \Delta\tau_u\le G/2$.
The rotations incurred by the shift in the DFT window are compensated for in~\eqref{eq:compensated_freq} by multiplying the $k$th subcarrier by~$e^{{j\pi G k}/{N}}$.

%

\section{Impact of Timing and Frequency Offsets}
\label{sec:analysis}



We now derive an expression for the signal-to-interference-noise-and-distortion ratio (SINDR) at the UEs  for the case $\abs{\Delta\tau_u} \le N + G/2$ and~$\abs{\Delta\varepsilon_u} < 1$.
The derived expression, which captures the impact of residual timing and frequency offsets in the presence of 1-bit DACs at the BS,  provides insights into how accurate a synchronization algorithm needs to be for the system to operate at a target SINDR.
Due to space constraints, we limit our analysis to frequency-flat channels (i.e., $L=1$) and to Nyquist-rate sampling DACs (i.e., $\setS = \{ 0, 1, \dots, N-1 \}$ and $\textit{OSR} = 1$). For this setup,
\begin{IEEEeqnarray}{rCl} \label{eq:y_freq_flat}
r_u[n] &=& e^{\frac{j2\pi{\Delta\varepsilon_u}n}{N}} \vech_u^T\quantize\lefto(\vecx[n + \Delta\tau_u]\right)  + \tilde{w}_u[n] \IEEEeqnarraynumspace
\end{IEEEeqnarray}
where $\tilde{w}_u[n] = e^{\frac{j2\pi{\varepsilon_u^\text{est}}n}{N}} w_u[n +\tau_u^\text{est}]\distas \jpg(0,N_0)$.


\subsection{Linearization using Bussgang's Theorem}

In what follows, we assume Gaussian signaling. According to Bussgang's theorem~\cite{bussgang52a}, when $\vecx[n] \distas \jpg(\veczero_{B}, \matC_\vecx)$ with $\matC_\vecx = \sum_{u=1}^U \vecp_u\vecp_u^H$, we can write~\eqref{eq:quantize} as
\begin{IEEEeqnarray}{rCl} \label{eq:quantizer_linearized}
	\quantize(\vecx[n]) &=& \matA\vecx[n] + \vece[n]
\end{IEEEeqnarray}
where the non-Gaussian noise term $\vece[n] \in \opC^B$ is uncorrelated with $\vecx[n]$ and the matrix $\matA \in \opR^{B \times B}$ is given by~\cite[Eq.~(15)]{jacobsson17d}
\begin{IEEEeqnarray}{rCl}\label{eq:bussgang_gain}
	\matA &=& \sqrt{\frac{2}{\pi B}	} \, \matD_\vecx^{-1/2}.
\end{IEEEeqnarray}
Here, $\matD_\vecx =  \text{diag}(\matC_{\vecx})$. 
Let $\matC_\vece$ denote the covariance of $\vece[n]$.
By using that $\vecx[n]$ and $\vece[n]$ are uncorrelated as well as Van Vleck's arcsine law~\cite{van-vleck66a}, we find that
\begin{IEEEeqnarray}{rCl} \label{eq:arcsine_mimo}
	\matC_{\vece}
	&=& \frac{2}{\pi B} \lefto( \arcsin \lefto( \matD_\vecx^{-1/2} \Re\{ \matC_{\vecx} \} \matD_\vecx^{-1/2}\right)\right. \nonumber\\ 
	&& + \lefto. j  \arcsin\lefto( \matD_\vecx^{-1/2} \Im\{ \matC_{\vecx} \}\matD_\vecx^{-1/2} \right)\right) - \matA\matC_\vecx\matA. \IEEEeqnarraynumspace 
\end{IEEEeqnarray}

\subsection{Time-Domain Received Signal}

To avoid ISI from adjacent OFDM symbols, the DFT window in~\eqref{eq:compensated_freq} should only contain  samples  from the $i$th~OFDM symbol.
However, when $\Delta\tau_u < -G/2$, the DFT window includes also samples from the $(i-1)$th OFDM symbol, whereas when $\Delta\tau_u > G/2$ the DFT window includes also samples from the $(i+1)$th OFDM symbol. 
We let
\begin{IEEEeqnarray}{rCl} \label{eq:psi}
\psi(\Delta\tau_u) =
\begin{cases}
	- \Delta\tau_u - G/2 , & \Delta\tau_u < -G/2 \\
	\Delta\tau_u - G/2, & \Delta\tau_u > G/2 \\
	0, & \text{otherwise}
\end{cases}  \IEEEeqnarraynumspace 
\end{IEEEeqnarray}
be the number of received samples from adjacent OFDM symbols in the $i$th DFT window.
With this definition, by assuming that $\hat\vecp_u[k] = \vecp_u$ for $k = 0, 1, \dots, N-1$ and by inserting \eqref{eq:x_m_time}, \eqref{eq:x_m_freq}, \eqref{eq:y_freq_flat}, and~\eqref{eq:quantizer_linearized} into~\eqref{eq:tilt}, we can write the time-domain signal received during the $i$th DFT window as%
\begin{IEEEeqnarray}{rCl}
r_u^{(i)}[n]
&=& e^{\frac{j2\pi{\Delta\varepsilon_u}(n + i(N+G)- G/2)}{N}} \sum_{v=1}^{U} \vech_u^T\matA\vecp_v z_{u,v}^{(i)}[n] \IEEEeqnarraynumspace  \nonumber\\
&& + \vech_u^T \tilde\vece^{(i)}[n] + \tilde{w}_u^{(i)}[n] 
\label{eq:rx_m_time}
\end{IEEEeqnarray}
where
\begin{IEEEeqnarray}{rCl}
z_{u,v}^{(i)}[n] 
&=& 
\begin{cases}  s_v^{(i-1)}[n + N - \psi(\Delta\tau_u)],  \\
\quad\!\! 0 \le n \le \psi(\Delta\tau_u) - 1, \ \! \Delta\tau_u < - G/2 \\  
s_v^{(i+1)}[n - G + \psi(\Delta\tau_u)], \\ 
\quad\!\! N -\psi(\Delta\tau_u) \le n \le N-1, \ \! \Delta\tau_u > G/2 \\
 s_v^{(i)}[n - G/2 + \Delta\tau_u], \\ 
\quad\!\! \text{otherwise}. 
\end{cases} \ \ \
\end{IEEEeqnarray}
%
%
In \eqref{eq:rx_m_time}, we have set $\tilde\vece^{(i)}[n] = e^{\frac{j2\pi{\Delta\varepsilon_u}(n + i(N+G) - G/2)}{N}} \vece[n + i(N+G) - G/2 + \Delta\tau_u]$ and $\tilde{w}_u^{(i)} = \tilde{w}_u[n + i(N+G) - G/2]$.


\subsection{Frequency-Domain Received Signal}


By inserting \eqref{eq:rx_m_time} into~\eqref{eq:compensated_freq}, one can show that the received signal on the $k$th subcarrier during the $i$th DFT window~is
\begin{IEEEeqnarray}{rCl} \label{eq:rx_m_freq}
\hat{r}_u^{(i)}[k]	
&=& \beta(\Delta\tau_u, \Delta\varepsilon_u) e^{j\phi_k^{(i)}(\Delta\tau_u, \Delta\varepsilon_u)} \vech_u^T\matA\vecp_u \hat{s}_u^{(i)}[k] \IEEEeqnarraynumspace \nonumber\\
&& + \hat{\imath}_u^{\,\text{isi}, (i)}[k]  + \hat{\imath}_u^{\,\text{ici}, (i)}[k] + \hat{\imath}_u^{\,\text{mui},(i)}[k] \nonumber\\ 
&& +  \vech_u^T \hat\vece^{(i)}[k] + \hat{w}_u^{(i)}[k] 
\end{IEEEeqnarray}
where
\begin{IEEEeqnarray}{rCl}
\beta(\Delta\tau_u, \Delta\varepsilon_u) &=& \frac{\sin\lefto({\pi\Delta\varepsilon_u(N-\psi(\Delta\tau_u))}/{N} \right)}{N \sin\lefto({\pi \Delta\varepsilon_u}/{N} \right)} \IEEEeqnarraynumspace  
\end{IEEEeqnarray}
and 
\begin{IEEEeqnarray}{rCl}
\!\phi_k^{(i)}(\Delta\tau_u, \Delta\varepsilon_u) &=& {2\pi\lefto(\Delta\tau_u k + \Delta\varepsilon_u(N+G)i\right)}/{N} \nonumber\\
&& - {\pi\Delta\varepsilon_u\psi(\Delta\tau_u) \sign(\Delta\tau_u \!-\! G/2)}/{N} \IEEEeqnarraynumspace \nonumber\\
&& + {\pi\Delta\varepsilon_u (N - G - 1)}/{N} 
\end{IEEEeqnarray}
are the attenuation and phase rotation caused by the residual STO and CFO.
In~\eqref{eq:rx_m_freq}, $\hat\vece^{(i)}[k]$ and $\hat{w}_u^{(i)}[k]$ are found by computing the DFT of $\tilde\vece^{(i)}[n]$ and $\tilde{w}_u^{(i)}[n]$, respectively.
Closed-form expressions for the ISI, ICI, and MU interference (MUI) $\hat{\imath}^{\,\text{isi},(i)}_u[k]$, $\hat{\imath}^{\,\text{ici},(i)}_u[k]$, and $\hat{\imath}^{\,\text{mui},(i)}_u[k]$, respectively, are found by computing the DFT of $\sum_{v=1}^{U} \vech_u^T\matA\vecp_v z_{u,v}^{(i)}[n]$, but are not provided here due to space constraints.

\subsection{Impact of Residual STO and CFO on the SINDR}

%
%
It can be shown that the five noise terms in~\eqref{eq:rx_m_freq} are uncorrelated with each other and with the transmitted symbol. 
Hence, we can write the SINDR for the $u$th UE as
\begin{IEEEeqnarray}{rCl} \label{eq:sindr}
	\textit{SINDR}_u &=& \frac{\abs{\beta(\Delta\tau_u, \Delta\varepsilon_u)}^2 \abs{\vech_u^T\matA\vecp_u}^2}{I_u^\text{isi} + I_u^\text{ici} + I_u^\text{mui} + \vech_u^T\matC_\vece\vech_u^* + N_0}.
\end{IEEEeqnarray}
Here, $I_u^\text{isi}$ is the power of the ISI term, $I_u^\text{ici}$ is the power of the ICI term, and $I_u^\text{mui}$ is the power of the MUI term.
By following steps similar to those in~\cite[App.~A and App.~B]{mostofi06a}, it can be shown that
\begin{IEEEeqnarray}{rCl}
I_u^\text{isi}
&=& \Ex{}{\big|\hat{\imath}^{\,\text{isi},(i)}_u[k]\big|^2} 
= \frac{\psi(\Delta\tau_u)}{N}\abs{\vech_u^T\matA\vecp_u}^2.
\end{IEEEeqnarray}
Furthermore,
\begin{IEEEeqnarray}{rCl}
I_u^\text{ici}
&=& \Ex{}{\big|\hat{\imath}^{\,\text{ici},(i)}_u[k]\big|^2} \nonumber\\
&=& \lefto( 1 - \abs{\beta(\Delta\tau_u, \Delta\varepsilon_u)}^2 - \frac{\psi(\Delta\tau_u)}{N} \right) \abs{\vech_u^T\matA\vecp_u}^2 \IEEEeqnarraynumspace
\end{IEEEeqnarray}
and
\begin{IEEEeqnarray}{rCl}
I_u^\text{mui} 
&=& \Ex{}{\big|\hat{\imath}^{\,\text{mui},(i)}_u[k]\big|^2} 
= \sum_{v=1, v \neq u}^U \abs{\vech_u^T \matA \vecp_{v}}^2.
\end{IEEEeqnarray}
Note that for the case $\Delta\tau_u = 0$ and $\Delta\varepsilon_u = 0$, the SINDR in~\eqref{eq:sindr} coincides with the expression given in~\cite[Eq.~(26)]{jacobsson17d}.

\subsection{Numerical Results}

\begin{figure}[!t]
\centering
\subfloat[SINDR versus residual STO for different values of CFO.]{\includegraphics[width = .9\columnwidth]{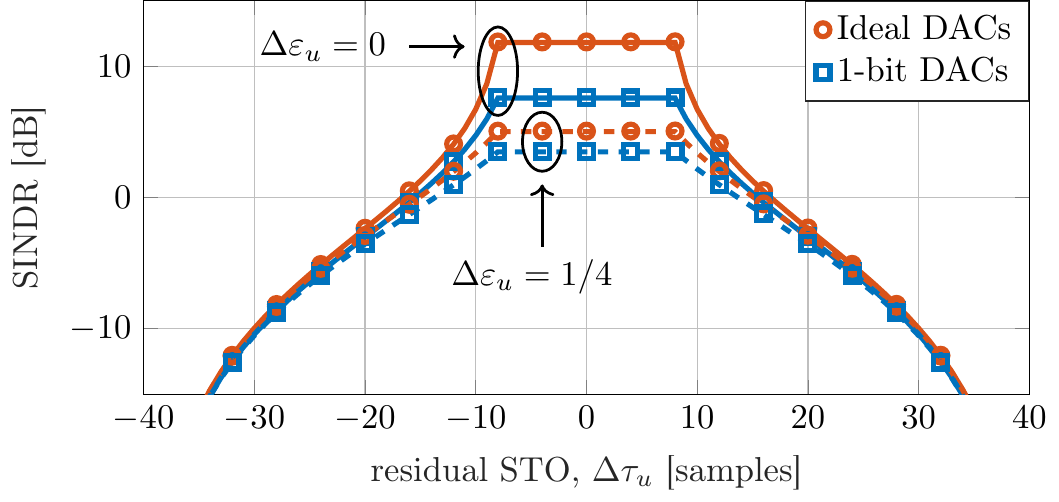}} \\ 
\subfloat[SINDR versus residual CFO for different values of STO.]{\includegraphics[width = .9\columnwidth]{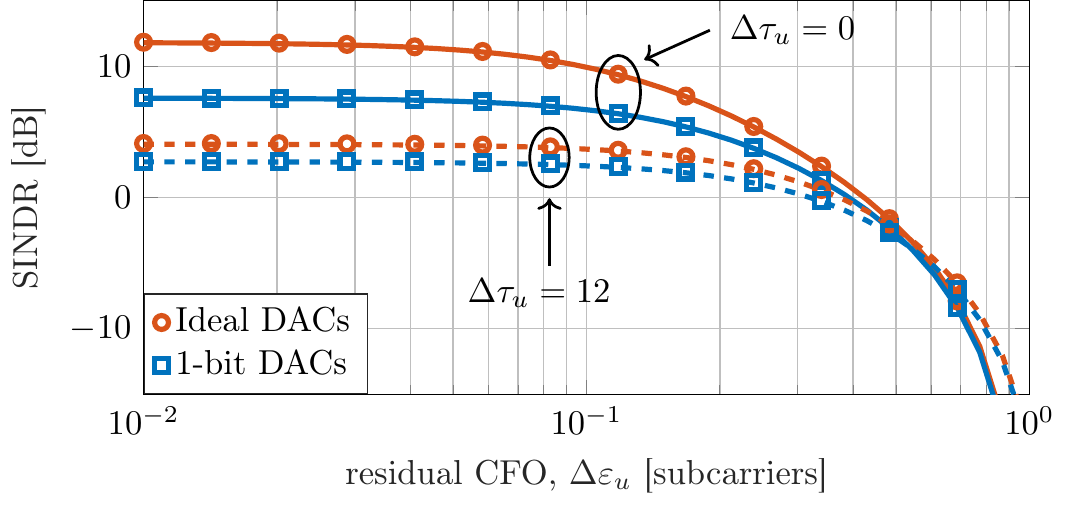}}
\caption{Impact of residual STO and CFO on the SINDR. We use ZF precoding, $N_0 = 0$\,dB, $B = 128$ antennas, $U = 8$ UEs, $N = S = 32$ subcarriers, and $G = 16$ samples. The solid and dashed lines correspond to analytical results and the markers correspond to simulation~results.}
\label{fig:sto_and_cfo}
\end{figure}

We verify our analytical results by means of numerical simulations.
We consider zero-forcing (ZF) precoding with perfect BS-side channel state information (CSI).
Furthermore, we assume that the effective channel gain $\beta(\Delta\tau_u, \Delta\varepsilon_u) e^{j\phi_k^{(i)}(\Delta\tau_u, \Delta\varepsilon_u)} \vech_u^T\matA\vecp_u$ in~\eqref{eq:rx_m_freq} is known at the $u$th UE.
In Section~\ref{sec:chest}, we discuss how such a knowledge can be acquired.
%
%
We fix $N_0 = 0$\,dB, $B=128$~antennas, $U = 8$~UEs, $N=S=32$ subcarriers, and a cyclic prefix of length $G = 16$~samples. 
In~\fref{fig:sto_and_cfo}, we plot the SINDR (averaged over $100$ channel realizations and over the UEs) versus the residual STO and CFO.
Note that the numerical simulations match our analytical results, which verifies the accuracy of the~analysis. 
One can see from~\fref{fig:sto_and_cfo} that the impact of the STO and the CFO on performance in the 1-bit-DAC case is similar to that in the infinite-resolution (quantization-free) case. 
In particular, we note that the CP provides some protection against timing errors as there is no loss in SINDR when $-G/2 \le \Delta\tau_u \le G/2$.
Increasing the length of the CP results in higher robustness against~timing~errors.

\section{Timing and Frequency Synchronization} \label{sec:synch}

Next, we show how the Schmidl-Cox algorithm proposed in~\cite{schmidl97a} can be used to establish timing and frequency synchronization in the 1-bit massive MU-MIMO-OFDM downlink. In this section, we consider the general case of frequency-selective channels and oversampling DACs.

\subsection{Estimation of STO and CFO using Schmidl-Cox Algorithm} \label{sec:estimation}

%
%
%
%
%
%


As in~\cite{schmidl97a}, we reserve the $0$th OFDM symbol for transmitting a preamble consisting of two identical sequences of  $N/2$ samples, such that $s_u^{(0)}[n + N/2] = s_u^{(0)}[n]$ for $n = 0,1,\dots, N/2-1$ and $u = 1,2,\dots, U$. 
%
%
It follows that $y_u[n] = z_u[n] + w_u[n]$ and $y_u[n + N/2] = e^{-j\pi{\varepsilon_u}} z_u[n] + w_u[n + N/2]$ for $ \tau_u \le  n \le \tau_u + N/2 -1$ and $u = 1,2,\dots, U$, where $z_u[n] =  e^{-\frac{j2\pi{\varepsilon_u}n}{N}} \sum_{\ell = 0}^{L-1}\vech_u^T[\ell] \quantize\lefto(\vecx[n-\ell]\right)$.
Note that, in the absence of AWGN, the two corresponding received preamble sequences, which include quantization noise and MUI, are identical except for a phase shift of $-\pi\varepsilon_u$ caused by the~CFO.
In particular, the presence of 1-bit DACs does not affect the symmetric structure of the preamble.
%
%
We use this property to compute an estimate of the STO $\tau_u$ at the $u$th UE as~\cite[Sec.~III-A]{minn00a}
\begin{IEEEeqnarray}{rCl} \label{eq:sto_est}
	\tau_u^\text{est} &=& \argmax_{\tau} \ \abs{\Gamma_u(\tau)}
\end{IEEEeqnarray}
where  $\Gamma_u(\tau) \in [0,1]$ is given by
\begin{IEEEeqnarray}{rCl} \label{eq:gamma}
	\Gamma_u(\tau)	 &=& \frac{1}{G+1} \sum_{n=-G}^0 \frac{\abs{P_u(n+\tau)}^2 }{R_u(n+\tau) }.
\end{IEEEeqnarray}
Here, $P_u(\tau) = \sum_{n=0}^{N/2-1} y_u[n+\tau] y_u^*[n + N/2 + \tau]$ and $R_u(\tau) = \frac{1}{2}\sum_{n=0}^{N-1} \abs{y_u[n + \tau]}^2$.
Next, we compute an estimate of the CFO $\varepsilon_u$ at the $u$th UE as follows~\cite{schmidl97a}:
\begin{IEEEeqnarray}{rCl} \label{eq:cfo_est}
	\varepsilon_u^\text{est} = \frac{1}{\pi}\arg\{ P_u(\tau_u^\text{est})\}.	
\end{IEEEeqnarray}
The estimator in~\eqref{eq:cfo_est} can be used whenever $\abs{\varepsilon_u} < 1$. The acquisition range can, however, be increased by transmitting an additional preamble symbol~(see, e.g.,~\cite{schmidl97a} for the details).

\subsection{Channel Estimation and Symbol Equalization}
\label{sec:chest}

Unless perfect synchronization has been achieved, residual STO and CFO cause an attenuation and a phase rotation that must be compensated for at the UEs.
We obtain an estimate of the effective channel gain, which includes the attenuation and the phase shift caused by residual STO and CFO, using least-squares (LS) channel estimation based on $P$ downlink training symbols as $\hat\alpha_u[k] = \frac{1}{P}\sum_{i = 1}^{P} \hat{r}_u^{(i)}[k] (\hat{s}_u^{(i)}[k])^*$.
Here, $\hat{s}_u^{(i)}[k]$ for $k \in \setS$ and $i = 1,2,\dots,P$ are the transmitted training symbols (known to the BS and to the UEs). 
Finally, for $k \in \setS$ and $i \notin \{ 0, 1, \dots, P \}$, we compute an estimate $\hat{s}_u^{\text{est}, (i)}[k]$ of $\hat{s}_u^{(i)}[k]$ as $\hat{s}_u^{\text{est}, (i)}[k] = {\hat\alpha^*_u[k]\hat{r}^{(i)}_u[k]}/{\abs{\hat\alpha_u[k]}^2}$.

\subsection{Numerical Results}

%
We consider an OFDM system with $N = 2048$ subcarriers, $S = 1200$ used subcarriers, and a cyclic prefix of length $G = 144$ samples. The OSR is $\textit{OSR} \approx 1.7$ and the used subcarriers are the first $600$ to the left and to the right of the DC subcarrier, i.e.,~$\setS = \{ 1, 2, \dots,S/2, N-S/2, N-S/2+1, \dots, N-1\}$.
In what follows, we fix $B = 128$ antennas, $U = 8$ UEs, and $L = 10$ taps. The STO and CFO are $\tau_u \in \setU(-N - G/2, N + G/2)$ and $\varepsilon_u \in \setU(-1,1)$ for $u = 1,2,\dots,U$.
We average all numerical simulations over $100$ random channel realizations and over the~UEs.
For each channel realization, we transmit $10$ OFDM symbols (excluding preamble and pilot symbols to estimate the effective channel gain). Again, we  assume that perfect CSI is available at the BS and that ZF precoding is~used.

\begin{figure}[!t]
\centering
\subfloat[RMSE of the STO estimate with the Schmidl-Cox algorithm.]{\includegraphics[width = .88\columnwidth]{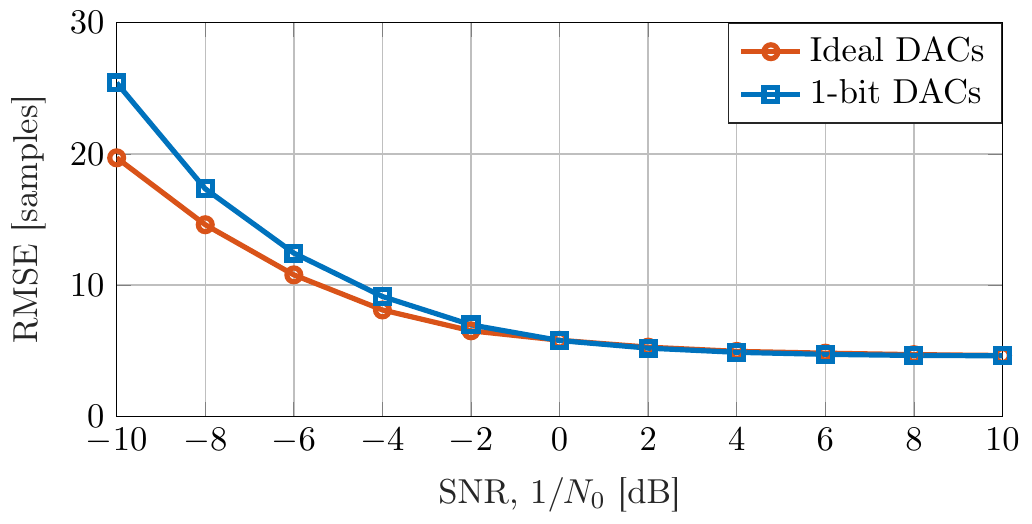}} \\ 
\subfloat[RMSE of the CFO estimate with the Schmidl-Cox algorithm.]{\includegraphics[width = .9\columnwidth]{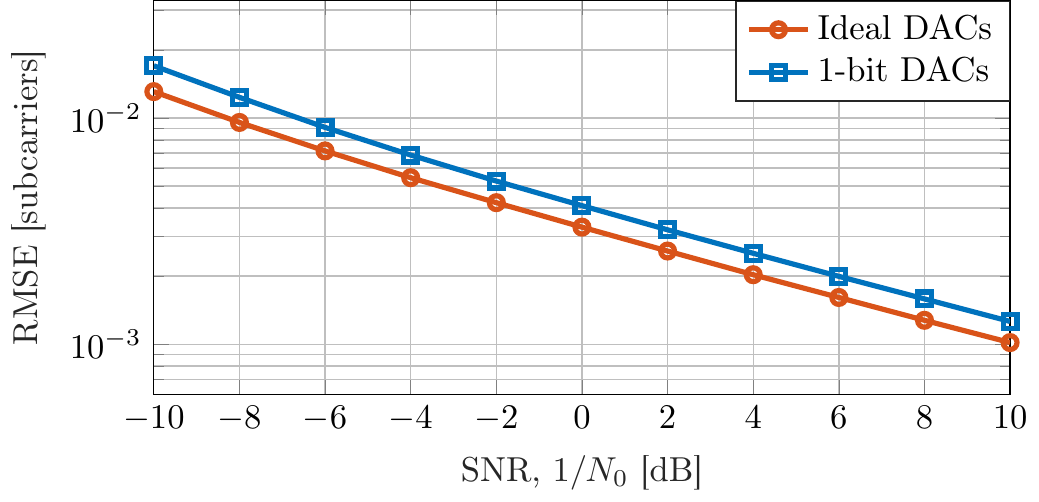}}
\caption{RMSE of the STO and CFO estimates with the Schmidl-Cox algorithm; ZF precoding, $B = 128$ antennas, $U = 8$ UEs, $N = 2048$ subcarriers, $S = 1200$ used subcarriers, $G = 144$ samples, and $L = 10$ taps.}
\label{fig:rmse}
\end{figure}

In~\fref{fig:rmse}, we show the root-mean-square error~(RMSE) of the timing and frequency estimators in~\eqref{eq:sto_est} and~\eqref{eq:cfo_est}, respectively. 
%
%
We note that, at high SNR, the RMSE of the STO estimate is well within acceptable limits (note that there is no ISI if $-63 \le \Delta\tau_u\le 72$ for $G=144$ and $L = 10$).
We further note that the RMSE of the CFO estimate with 1-bit DACs is only slightly higher than that in the infinite-resolution case. The SNR gap is approximately $2$\,dB, which is in accordance with the factor $\sqrt{2/\pi}$ in~\eqref{eq:bussgang_gain}.

\begin{figure}[!t]
	\centering
	\includegraphics[width = .9\columnwidth]{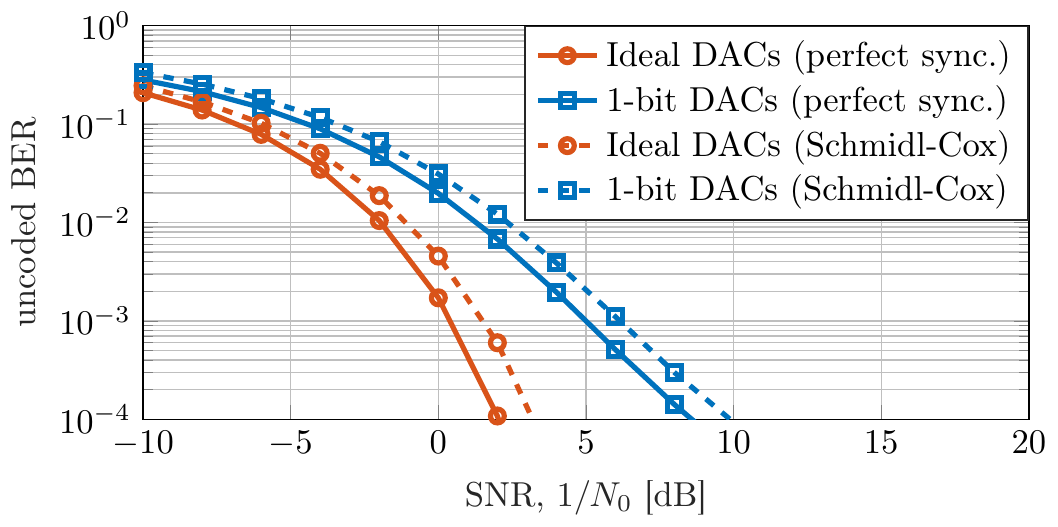}
	\caption{Uncoded BER with QPSK and with the Schmidl-Cox synchronization algorithm followed by LS channel estimation; ZF precoding, $B = 128$ antennas, $U = 8$ UEs, $N = 2048$ subcarriers, $S = 1200$ used subcarriers, $G = 144$~samples, $L = 10$~taps, and $P = 1$ training symbols.}
	\label{fig:ber}
\end{figure}

In~\fref{fig:ber}, we show the uncoded BER with QPSK symbols. 
%
%
Here, to minimize training overhead, we transmit only one downlink training symbol~(i.e., $P=1$). 
We observe from the figure that Schmidl-Cox synchronization followed by LS channel estimation incurs a moderate SNR loss compared to the case of perfect timing and frequency synchronization (i.e., when the $u$th UE knows \emph{a priori} the realizations of $\tau_u$ and~$\varepsilon_u$). 
%
%
This implies that the 1-bit massive MU-MIMO-OFDM downlink is, to some extent, resilient towards timing and frequency offsets.

\section{Conclusions} \label{sec:conclusions}

We have evaluated the performance of the well-known Schmidl-Cox algorithm for achieving timing and frequency synchronization over frequency-selective channels in the 1-bit massive MU-MIMO-OFDM downlink. For the case of frequency-flat channels, we have further characterized analytically the joint impact of residual STO and CFO on the SINDR at the UEs.

Our results suggest that sufficiently accurate timing and frequency synchronization can be achieved despite the quantization artifacts introduced by the 1-bit DACs by simply using algorithms such as the Schmidl-Cox algorithm, which have been developed for the infinite-resolution~case.   

An extension of the analysis reported in~\fref{sec:analysis} to the frequency-selective channels will be presented in an upcoming extension of this paper. 
%

%
%
%
%
%
%
%

%

\begin{spacing}{.963}
\bibliographystyle{IEEEtran}
\bibliography{IEEEabrv,confs-jrnls,publishers,svenbib}		
\end{spacing}

\end{document}

%% file: vmr-symbols-vecbold.tex
\usepackage{amssymb}
\usepackage{amsfonts}
\usepackage{mathrsfs}
\usepackage{xspace}
\usepackage{bm}
\usepackage{upgreek}

\newcommand{\safemath}[2]{\newcommand{#1}{\ensuremath{#2}\xspace}}

\safemath{\bma}{\mathbf{a}}
\safemath{\bmb}{\mathbf{b}}
\safemath{\bmc}{\mathbf{c}}
\safemath{\bmd}{\mathbf{d}}
\safemath{\bme}{\mathbf{e}}
\safemath{\bmf}{\mathbf{f}}
\safemath{\bmg}{\mathbf{g}}
\safemath{\bmh}{\mathbf{h}}
\safemath{\bmi}{\mathbf{i}}
\safemath{\bmj}{\mathbf{j}}
\safemath{\bmk}{\mathbf{k}}
\safemath{\bml}{\mathbf{l}}
\safemath{\bmm}{\mathbf{m}}
\safemath{\bmn}{\mathbf{n}}
\safemath{\bmo}{\mathbf{o}}
\safemath{\bmp}{\mathbf{p}}
\safemath{\bmq}{\mathbf{q}}
\safemath{\bmr}{\mathbf{r}}
\safemath{\bms}{\mathbf{s}}
\safemath{\bmt}{\mathbf{t}}
\safemath{\bmu}{\mathbf{u}}
\safemath{\bmv}{\mathbf{v}}
\safemath{\bmw}{\mathbf{w}}
\safemath{\bmx}{\mathbf{x}}
\safemath{\bmy}{\mathbf{y}}
\safemath{\bmz}{\mathbf{z}}
\safemath{\bmzero}{\mathbf{0}}
\safemath{\bmone}{\mathbf{1}}

\bmdefine{\biad}{a}
\bmdefine{\bibd}{b}
\bmdefine{\bicd}{c}
\bmdefine{\bidd}{d}
\bmdefine{\bied}{e}
\bmdefine{\bifd}{f}
\bmdefine{\bigd}{g}
\bmdefine{\bihd}{h}
\bmdefine{\biid}{i}
\bmdefine{\bijd}{j}
\bmdefine{\bikd}{k}
\bmdefine{\bild}{l}
\bmdefine{\bimd}{m}
\bmdefine{\bind}{n}
\bmdefine{\biod}{o}
\bmdefine{\bipd}{p}
\bmdefine{\biqd}{q}
\bmdefine{\bird}{r}
\bmdefine{\bisd}{s}
\bmdefine{\bitd}{t}
\bmdefine{\biud}{u}
\bmdefine{\bivd}{v}
\bmdefine{\biwd}{w}
\bmdefine{\bixd}{x}
\bmdefine{\biyd}{y}
\bmdefine{\bizd}{z}

\bmdefine{\bixid}{\xi}
\bmdefine{\bilambdad}{\lambda}
\bmdefine{\bimud}{\mu}
\bmdefine{\bithetad}{\theta}
\bmdefine{\biphid}{\phi}
\bmdefine{\bideltad}{\delta}

\safemath{\bmia}{\biad}
\safemath{\bmib}{\bibd}
\safemath{\bmic}{\bicd}
\safemath{\bmid}{\bidd}
\safemath{\bmie}{\bied}
\safemath{\bmif}{\bifd}
\safemath{\bmig}{\bigd}
\safemath{\bmih}{\bihd}
\safemath{\bmii}{\biid}
\safemath{\bmij}{\bijd}
\safemath{\bmik}{\bikd}
\safemath{\bmil}{\bild}
\safemath{\bmim}{\bimd}
\safemath{\bmin}{\bind}
\safemath{\bmio}{\biod}
\safemath{\bmip}{\bipd}
\safemath{\bmiq}{\biqd}
\safemath{\bmir}{\bird}
\safemath{\bmis}{\bisd}
\safemath{\bmit}{\bitd}
\safemath{\bmiu}{\biud}
\safemath{\bmiv}{\bivd}
\safemath{\bmiw}{\biwd}
\safemath{\bmix}{\bixd}
\safemath{\bmiy}{\biyd}
\safemath{\bmiz}{\bizd}

\safemath{\bmxi}{\bixid}
\safemath{\bmlambda}{\bilambdad}
\safemath{\bmmu}{\bimud}
\safemath{\bmtheta}{\bithetad}
\safemath{\bmphi}{\biphid}
\safemath{\bmdelta}{\bideltad}

\safemath{\bA}{\mathbf{A}}
\safemath{\bB}{\mathbf{B}}
\safemath{\bC}{\mathbf{C}}
\safemath{\bD}{\mathbf{D}}
\safemath{\bE}{\mathbf{E}}
\safemath{\bF}{\mathbf{F}}
\safemath{\bG}{\mathbf{G}}
\safemath{\bH}{\mathbf{H}}
\safemath{\bI}{\mathbf{I}}
\safemath{\bJ}{\mathbf{J}}
\safemath{\bK}{\mathbf{K}}
\safemath{\bL}{\mathbf{L}}
\safemath{\bM}{\mathbf{M}}
\safemath{\bN}{\mathbf{N}}
\safemath{\bO}{\mathbf{O}}
\safemath{\bP}{\mathbf{P}}
\safemath{\bQ}{\mathbf{Q}}
\safemath{\bR}{\mathbf{R}}
\safemath{\bS}{\mathbf{S}}
\safemath{\bT}{\mathbf{T}}
\safemath{\bU}{\mathbf{U}}
\safemath{\bV}{\mathbf{V}}
\safemath{\bW}{\mathbf{W}}
\safemath{\bX}{\mathbf{X}}
\safemath{\bY}{\mathbf{Y}}
\safemath{\bZ}{\mathbf{Z}}

\safemath{\bZero}{\mathbf{0}}
\safemath{\bOne}{\mathbf{1}}
\safemath{\bDelta}{\mathbf{\Delta}}
\safemath{\bLambda}{\mathbf{\UpLambda}}
\safemath{\bPhi}{\mathbf{\Upphi}}
\safemath{\bSigma}{\mathbf{\Upsigma}}
\safemath{\bOmega}{\mathbf{\Upomega}}
\safemath{\bTheta}{\mathbf{\Uptheta}}

\bmdefine{\biAd}{A}
\bmdefine{\biBd}{B}
\bmdefine{\biCd}{C}
\bmdefine{\biDd}{D}
\bmdefine{\biEd}{E}
\bmdefine{\biFd}{F}
\bmdefine{\biGd}{G}
\bmdefine{\biHd}{H}
\bmdefine{\biId}{I}
\bmdefine{\biJd}{J}
\bmdefine{\biKd}{K}
\bmdefine{\biLd}{L}
\bmdefine{\biMd}{M}
\bmdefine{\biOd}{N}
\bmdefine{\biPd}{O}
\bmdefine{\biQd}{P}
\bmdefine{\biRd}{R}
\bmdefine{\biSd}{S}
\bmdefine{\biTd}{T}
\bmdefine{\biUd}{U}
\bmdefine{\biVd}{V}
\bmdefine{\biWd}{W}
\bmdefine{\biXd}{X}
\bmdefine{\biYd}{Y}
\bmdefine{\biZd}{Z}

\bmdefine{\biDelta}{\Delta}
\bmdefine{\biLambda}{\Lambda}
\bmdefine{\biPhi}{\Phi}
\bmdefine{\biSigma}{\Sigma}
\bmdefine{\biOmega}{\Omega}
\bmdefine{\biTheta}{\Theta}

\safemath{\bimA}{\biAd}
\safemath{\bimB}{\biBd}
\safemath{\bimC}{\biCd}
\safemath{\bimD}{\biDd}
\safemath{\bimE}{\biEd}
\safemath{\bimF}{\biFd}
\safemath{\bimG}{\biGd}
\safemath{\bimH}{\biHd}
\safemath{\bimI}{\biId}
\safemath{\bimJ}{\biJd}
\safemath{\bimK}{\biKd}
\safemath{\bimL}{\biLd}
\safemath{\bimM}{\biMd}
\safemath{\bimN}{\biNd}
\safemath{\bimO}{\biOd}
\safemath{\bimP}{\biPd}
\safemath{\bimQ}{\biQd}
\safemath{\bimR}{\biRd}
\safemath{\bimS}{\biSd}
\safemath{\bimT}{\biTd}
\safemath{\bimU}{\biUd}
\safemath{\bimV}{\biVd}
\safemath{\bimW}{\biWd}
\safemath{\bimX}{\biXd}
\safemath{\bimY}{\biYd}
\safemath{\bimZ}{\biZd}

\safemath{\bimDelta}{\biDelta}
\safemath{\bimLambda}{\biLambda}
\safemath{\bimPhi}{\biPhi}
\safemath{\bimSigma}{\biSigma}
\safemath{\bimOmega}{\biOmega}
\safemath{\bimTheta}{\biTheta}

\safemath{\setA}{\mathcal{A}}
\safemath{\setB}{\mathcal{B}}
\safemath{\setC}{\mathcal{C}}
\safemath{\setD}{\mathcal{D}}
\safemath{\setE}{\mathcal{E}}
\safemath{\setF}{\mathcal{F}}
\safemath{\setG}{\mathcal{G}}
\safemath{\setH}{\mathcal{H}}
\safemath{\setI}{\mathcal{I}}
\safemath{\setJ}{\mathcal{J}}
\safemath{\setK}{\mathcal{K}}
\safemath{\setL}{\mathcal{L}}
\safemath{\setM}{\mathcal{M}}
\safemath{\setN}{\mathcal{N}}
\safemath{\setO}{\mathcal{O}}
\safemath{\setP}{\mathcal{P}}
\safemath{\setQ}{\mathcal{Q}}
\safemath{\setR}{\mathcal{R}}
\safemath{\setS}{\mathcal{S}}
\safemath{\setT}{\mathcal{T}}
\safemath{\setU}{\mathcal{U}}
\safemath{\setV}{\mathcal{V}}
\safemath{\setW}{\mathcal{W}}
\safemath{\setX}{\mathcal{X}}
\safemath{\setY}{\mathcal{Y}}
\safemath{\setZ}{\mathcal{Z}}
\safemath{\emptySet}{\varnothing}

\safemath{\colA}{\mathscr{A}}
\safemath{\colB}{\mathscr{B}}
\safemath{\colC}{\mathscr{C}}
\safemath{\colD}{\mathscr{D}}
\safemath{\colE}{\mathscr{E}}
\safemath{\colF}{\mathscr{F}}
\safemath{\colG}{\mathscr{G}}
\safemath{\colH}{\mathscr{H}}
\safemath{\colI}{\mathscr{I}}
\safemath{\colJ}{\mathscr{J}}
\safemath{\colK}{\mathscr{K}}
\safemath{\colL}{\mathscr{L}}
\safemath{\colM}{\mathscr{M}}
\safemath{\colN}{\mathscr{N}}
\safemath{\colO}{\mathscr{O}}
\safemath{\colP}{\mathscr{P}}
\safemath{\colQ}{\mathscr{Q}}
\safemath{\colR}{\mathscr{R}}
\safemath{\colS}{\mathscr{S}}
\safemath{\colT}{\mathscr{T}}
\safemath{\colU}{\mathscr{U}}
\safemath{\colV}{\mathscr{V}}
\safemath{\colW}{\mathscr{W}}
\safemath{\colX}{\mathscr{X}}
\safemath{\colY}{\mathscr{Y}}
\safemath{\colZ}{\mathscr{Z}}

\safemath{\opA}{\mathbb{A}}
\safemath{\opB}{\mathbb{B}}
\safemath{\opC}{\mathbb{C}}
\safemath{\opD}{\mathbb{D}}
\safemath{\opE}{\mathbb{E}}
\safemath{\opF}{\mathbb{F}}
\safemath{\opG}{\mathbb{G}}
\safemath{\opH}{\mathbb{H}}
\safemath{\opI}{\mathbb{I}}
\safemath{\opJ}{\mathbb{J}}
\safemath{\opK}{\mathbb{K}}
\safemath{\opL}{\mathbb{L}}
\safemath{\opM}{\mathbb{M}}
\safemath{\opN}{\mathbb{N}}
\safemath{\opO}{\mathbb{O}}
\safemath{\opP}{\mathbb{P}}
\safemath{\opQ}{\mathbb{Q}}
\safemath{\opR}{\mathbb{R}}
\safemath{\opS}{\mathbb{S}}
\safemath{\opT}{\mathbb{T}}
\safemath{\opU}{\mathbb{U}}
\safemath{\opV}{\mathbb{V}}
\safemath{\opW}{\mathbb{W}}
\safemath{\opX}{\mathbb{X}}
\safemath{\opY}{\mathbb{Y}}
\safemath{\opZ}{\mathbb{Z}}
\safemath{\opZero}{\mathbb{O}}
\safemath{\identityop}{\opI}

\safemath{\veca}{\bma}
\safemath{\vecb}{\bmb}
\safemath{\vecc}{\bmc}
\safemath{\vecd}{\bmd}
\safemath{\vece}{\bme}
\safemath{\vecf}{\bmf}
\safemath{\vecg}{\bmg}
\safemath{\vech}{\bmh}
\safemath{\veci}{\bmi}
\safemath{\vecj}{\bmj}
\safemath{\veck}{\bmk}
\safemath{\vecl}{\bml}
\safemath{\vecm}{\bmm}
\safemath{\vecn}{\bmn}
\safemath{\veco}{\bmo}
\safemath{\vecp}{\bmp}
\safemath{\vecq}{\bmq}
\safemath{\vecr}{\bmr}
\safemath{\vecs}{\bms}
\safemath{\vect}{\bmt}
\safemath{\vecu}{\bmu}
\safemath{\vecv}{\bmv}
\safemath{\vecw}{\bmw}
\safemath{\vecx}{\bmx}
\safemath{\vecy}{\bmy}
\safemath{\vecz}{\bmz}

\safemath{\veczero}{\bmzero}
\safemath{\vecone}{\bmone}
\safemath{\vecxi}{\bmxi}
\safemath{\veclambda}{\bmlambda}
\safemath{\vecmu}{\bmmu}
\safemath{\vectheta}{\bmtheta}
\safemath{\vecphi}{\bmphi}
\safemath{\vecdelta}{\bmdelta}

\safemath{\matA}{\bA}
\safemath{\matB}{\bB}
\safemath{\matC}{\bC}
\safemath{\matD}{\bD}
\safemath{\matE}{\bE}
\safemath{\matF}{\bF}
\safemath{\matG}{\bG}
\safemath{\matH}{\bH}
\safemath{\matI}{\bI}
\safemath{\matJ}{\bJ}
\safemath{\matK}{\bK}
\safemath{\matL}{\bL}
\safemath{\matM}{\bM}
\safemath{\matN}{\bN}
\safemath{\matO}{\bO}
\safemath{\matP}{\bP}
\safemath{\matQ}{\bQ}
\safemath{\matR}{\bR}
\safemath{\matS}{\bS}
\safemath{\matT}{\bT}
\safemath{\matU}{\bU}
\safemath{\matV}{\bV}
\safemath{\matW}{\bW}
\safemath{\matX}{\bX}
\safemath{\matY}{\bY}
\safemath{\matZ}{\bZ}
\safemath{\matzero}{\bmzero}

\safemath{\matDelta}{\bDelta}
\safemath{\matLambda}{\bLambda}
\safemath{\matPhi}{\bPhi}
\safemath{\matSigma}{\bSigma}
\safemath{\matOmega}{\bOmega}
\safemath{\matTheta}{\bTheta}

\safemath{\matidentity}{\matI}
\safemath{\matone}{\matO}

\safemath{\rnda}{A}
\safemath{\rndb}{B}
\safemath{\rndc}{C}
\safemath{\rndd}{D}
\safemath{\rnde}{E}
\safemath{\rndf}{F}
\safemath{\rndg}{G}
\safemath{\rndh}{H}
\safemath{\rndi}{I}
\safemath{\rndj}{J}
\safemath{\rndk}{K}
\safemath{\rndl}{L}
\safemath{\rndm}{M}
\safemath{\rndn}{N}
\safemath{\rndo}{O}
\safemath{\rndp}{P}
\safemath{\rndq}{Q}
\safemath{\rndr}{R}
\safemath{\rnds}{S}
\safemath{\rndt}{T}
\safemath{\rndu}{U}
\safemath{\rndv}{V}
\safemath{\rndw}{W}
\safemath{\rndx}{X}
\safemath{\rndy}{Y}
\safemath{\rndz}{Z}

\safemath{\rveca}{\bimA}
\safemath{\rvecb}{\bimB}
\safemath{\rvecc}{\bimC}
\safemath{\rvecd}{\bimD}
\safemath{\rvece}{\bimE}
\safemath{\rvecf}{\bimF}
\safemath{\rvecg}{\bimG}
\safemath{\rvech}{\bimH}
\safemath{\rveci}{\bimI}
\safemath{\rvecj}{\bimJ}
\safemath{\rveck}{\bimK}
\safemath{\rvecl}{\bimL}
\safemath{\rvecm}{\bimM}
\safemath{\rvecn}{\bimN}
\safemath{\rveco}{\bomO}
\safemath{\rvecp}{\bimP}
\safemath{\rvecq}{\bimQ}
\safemath{\rvecr}{\bimR}
\safemath{\rvecs}{\bimS}
\safemath{\rvect}{\bimT}
\safemath{\rvecu}{\bimU}
\safemath{\rvecv}{\bimV}
\safemath{\rvecw}{\bimW}
\safemath{\rvecx}{\bimX}
\safemath{\rvecy}{\bimY}
\safemath{\rvecz}{\bimZ}

\safemath{\rvecxi}{\bmxi}
\safemath{\rveclambda}{\bmlambda}
\safemath{\rvecmu}{\bmmu}
\safemath{\rvectheta}{\bmtheta}
\safemath{\rvecphi}{\bmphi}

\safemath{\rmatA}{\bimA}
\safemath{\rmatB}{\bimB}
\safemath{\rmatC}{\bimC}
\safemath{\rmatD}{\bimD}
\safemath{\rmatE}{\bimE}
\safemath{\rmatF}{\bimF}
\safemath{\rmatG}{\bimG}
\safemath{\rmatH}{\bimH}
\safemath{\rmatI}{\bimI}
\safemath{\rmatJ}{\bimJ}
\safemath{\rmatK}{\bimK}
\safemath{\rmatL}{\bimL}
\safemath{\rmatM}{\bimM}
\safemath{\rmatN}{\bimN}
\safemath{\rmatO}{\bimO}
\safemath{\rmatP}{\bimP}
\safemath{\rmatQ}{\bimQ}
\safemath{\rmatR}{\bimR}
\safemath{\rmatS}{\bimS}
\safemath{\rmatT}{\bimT}
\safemath{\rmatU}{\bimU}
\safemath{\rmatV}{\bimV}
\safemath{\rmatW}{\bimW}
\safemath{\rmatX}{\bimX}
\safemath{\rmatY}{\bimY}
\safemath{\rmatZ}{\bimZ}

\safemath{\rmatDelta}{\bimDelta}
\safemath{\rmatLambda}{\bimLambda}
\safemath{\rmatPhi}{\bimPhi}
\safemath{\rmatSigma}{\bimSigma}
\safemath{\rmatOmega}{\bimOmega}
\safemath{\rmatTheta}{\bimTheta}

%% file: standard-macros.tex
\usepackage{amssymb}
\usepackage{amsfonts}
\usepackage{mathrsfs}
\usepackage{xspace}
\usepackage{bm}
\usepackage{fancyref}
\usepackage{textcomp}

\usepackage{multirow}
\usepackage{stmaryrd}

\newenvironment{textbmatrix}{	\setlength{\arraycolsep}{2.5pt}%
								\big[\begin{matrix}}{\end{matrix}\big]%
								\raisebox{0.08ex}{\vphantom{M}}}

\def\be{\begin{equation}}
\def\ee{\end{equation}}
\def\een{\nonumber \end{equation}}
\def\mat{\begin{bmatrix}}
\def\emat{\end{bmatrix}}
\def\btm{\begin{textbmatrix}}
\def\etm{\end{textbmatrix}}

\def\ba#1\ea{\begin{align}#1\end{align}}
\def\bas#1\eas{\begin{align*}#1\end{align*}}
\def\bs#1\es{\begin{split}#1\end{split}} 
\def\bg#1\eg{\begin{gather}#1\end{gather}}
\def\bml#1\eml{\begin{multline}#1\end{multline}}
\def\bi#1\ei{\begin{itemize}#1\end{itemize}}

\newcommand{\subtext}[1]{\text{\fontfamily{cmr}\fontshape{n}\fontseries{m}\selectfont{}#1}}
\newcommand{\lefto}{\mathopen{}\left}

\DeclareMathOperator{\sign}{sgn}			
\DeclareMathOperator*{\argmax}{arg\;max}		
\DeclareMathOperator{\Exop}{\opE}			
\DeclareMathOperator{\Varop}{\mathrm{Var}}
\DeclareMathOperator{\Covop}{\mathrm{Cov}}
\DeclareMathOperator{\rot}{\nabla\times}		

\newcommand{\Ex}[2]{\ensuremath{\Exop_{#1}\lefto[#2\right]}} 	
\newcommand{\abs}[1]{\lefto\lvert#1\right\rvert}		

\newcommand{\vecnorm}[1]{\lefto\lVert#1\right\rVert}		
\newcommand{\opnorm}[1]{\lVert#1\rVert}		

\safemath{\dirac}{\delta}					
\safemath{\krond}{\dirac}					

\safemath{\upto}{\uparrow}
\safemath{\downto}{\downarrow}
\safemath{\iu}{j}							
\safemath{\ev}{\lambda}						
\safemath{\hilseqspace}{l^{2}}				
\newcommand{\banachfunspace}[1]{\setL^{#1}}	
\safemath{\hilfunspace}{\banachfunspace{2}}	

\safemath{\SNR}{\textsf{SNR}} 				
\safemath{\PAR}{\textsf{PAR}} 				
\safemath{\No}{N_0}							
\safemath{\Es}{E_s}							
\safemath{\Eb}{E_b}							
\safemath{\EbNo}{\frac{\Eb}{\No}}
\safemath{\EsNo}{\frac{\Es}{\No}}

\DeclareMathOperator{\CHop}{\ensuremath{\opH}} 
\safemath{\tvir}{\rndh_{\CHop}}				
\safemath{\tvtf}{\rndl_{\CHop}}				
\safemath{\spf}{\rnds_{\CHop}}				
\safemath{\bff}{H_{\CHop}}					

\safemath{\ircf}{r_{h}}						
\safemath{\tftvcf}{r_{s}}					
\safemath{\tfcf}{r_{l}}						
\safemath{\bfcf}{r_{H}}						

\safemath{\tcorr}{c_h}						
\safemath{\scf}{c_{s}}						
\safemath{\tfcorr}{c_{l}}					
\safemath{\fcorr}{c_{H}}						

\safemath{\mi}{I}							
\safemath{\capacity}{C}						

\safemath{\normal}{\mathcal{N}}			
\safemath{\jpg}{\mathcal{CN}}			
\safemath{\mchain}{\leftrightarrow}		
\newcommand{\AIC}{\ensuremath{\text{AIC}}} 

\safemath{\dB}{\,\mathrm{dB}}
\safemath{\dBm}{\,\mathrm{dBm}}
\safemath{\Hz}{\,\mathrm{Hz}}
\safemath{\kHz}{\,\mathrm{kHz}}
\safemath{\MHz}{\,\mathrm{MHz}}
\safemath{\GHz}{\,\mathrm{GHz}}
\safemath{\s}{\,\mathrm{s}}
\safemath{\ms}{\,\mathrm{ms}}
\safemath{\mus}{\,\mathrm{\text{\textmu}s}}
\safemath{\ns}{\,\mathrm{ns}}
\safemath{\ps}{\,\mathrm{ps}}
\safemath{\meter}{\,\mathrm{m}}
\safemath{\mm}{\,\mathrm{mm}}
\safemath{\cm}{\,\mathrm{cm}}
\safemath{\m}{\,\mathrm{m}}
\safemath{\W}{\,\mathrm{W}}
\safemath{\mW}{\, \mathrm{mW}}
\safemath{\J}{\,\mathrm{J}}
\safemath{\K}{\,\mathrm{K}}
\safemath{\bit}{\,\mathrm{bit}}
\safemath{\nat}{\,\mathrm{nat}}

\safemath{\define}{\triangleq}			

\safemath{\equivalent}{\sim}
\safemath{\distas}{\sim}					
\safemath{\sdiff}{\Delta}				

\safemath{\reals}{\mathbb{R}}
\safemath{\positivereals}{\reals_{+}}
\safemath{\integers}{\mathbb{Z}}
\safemath{\posint}{\integers_{+}}
\safemath{\naturals}{\mathbb{N}}
\safemath{\posnaturals}{\naturals_{+}}
\safemath{\complexset}{\mathbb{C}}
\safemath{\rationals}{\mathbb{Q}}

\newcommand*{\fancyrefapplabelprefix}{app}		
\newcommand*{\fancyrefthmlabelprefix}{thm}		
\newcommand*{\fancyreflemlabelprefix}{lem}		
\newcommand*{\fancyrefcorlabelprefix}{cor}		
\newcommand*{\fancyrefdeflabelprefix}{def}		
\newcommand*{\fancyrefproplabelprefix}{prop}	
\newcommand*{\fancyrefobslabelprefix}{obs}		
\newcommand*{\fancyrefalglabelprefix}{alg}		
\newcommand*{\fancyrefasmlabelprefix}{asm}	    
\newcommand*{\fancyreftbllabelprefix}{tab}	 

\frefformat{vario}{\fancyrefseclabelprefix}{Section~#1}
\frefformat{vario}{\fancyrefthmlabelprefix}{Theorem~#1}
\frefformat{vario}{\fancyreflemlabelprefix}{Lemma~#1}
\frefformat{vario}{\fancyrefcorlabelprefix}{Corollary~#1}
\frefformat{vario}{\fancyrefdeflabelprefix}{Definition~#1}
\frefformat{vario}{\fancyrefobslabelprefix}{Observation~#1}
\frefformat{vario}{\fancyrefasmlabelprefix}{Assumption~#1}
\frefformat{vario}{\fancyreffiglabelprefix}{Fig.~#1}
\frefformat{vario}{\fancyrefapplabelprefix}{Appendix~#1} 
\frefformat{vario}{\fancyrefproplabelprefix}{Proposition~#1}
\frefformat{vario}{\fancyrefalglabelprefix}{Algorithm~#1}
\frefformat{vario}{\fancyreftbllabelprefix}{Table~#1}
\frefformat{vario}{\fancyrefeqlabelprefix}{(#1)}

%% file: defs.tex
\newcommand{\marginnote}[1]{\marginpar{\framebox{\framebox{#1}}}}
\safemath{\dictab}{[\,\dicta\,\,\dictb\,]}

\safemath{\ysig}{\bmy}
\safemath{\ysighat}{\hat{\ysig}}
\safemath{\ysigdim}{M}
\safemath{\xsig}{\bmx}
\safemath{\xsigdim}{N}
\safemath{\nx}{n_x}
\safemath{\zsig}{\bmz}
\safemath{\zsigdim}{\ysigdim}
\safemath{\rsig}{\bmr}
\safemath{\Adict}{\bA}
\safemath{\Adicttilde}{\widetilde{\Adict}}
\safemath{\Adictdim}{\outputdim\times\xsigdim}
\safemath{\avec}{\bma}
\safemath{\avectilde}{\tilde{\avec}}
\safemath{\Bdict}{\bB}
\safemath{\Bdicttilde}{\widetilde{\Bdict}}
\safemath{\Cdict}{\bC}
\safemath{\cvec}{\bmc}
\safemath{\Ddict}{\bD}
\safemath{\Ddictdim}{\ysigdim\times\xsigdim}
\safemath{\dvec}{\bmd}
\safemath{\Ddicttilde}{\widetilde{\bD}}
\safemath{\Bonb}{\bB}
\safemath{\bvec}{\bmb}
\safemath{\Bonbdim}{\ysigdim\times\ysigdim}
\safemath{\noise}{\bmn}
\safemath{\noisedim}{\ysigim}
\safemath{\err}{\bme}
\safemath{\errdim}{\ysigdim}
\safemath{\errset}{\setE}
\safemath{\nerr}{n_e}
\safemath{\delop}{\bP_\errset}
\safemath{\delopc}{\bP_{{\errset}^c}}

\safemath{\cplxi}{\imath}
\safemath{\cplxj}{\jmath}

\safemath{\dict}{\matD}
\safemath{\inputdim}{N}		
\safemath{\outputdim}{M}		
\safemath{\sparsity}{S}	
\safemath{\inputdimA}{{N_a}}	
\safemath{\inputdimB}{{N_b}}	
\safemath{\elemA}{{n_a}}	
\safemath{\elemB}{{n_b}}	
\safemath{\resA}{\matR_a}	
\safemath{\resB}{\matR_b}	
\safemath{\subD}{\matS} 
\safemath{\subA}{\matS_a} 
\safemath{\subB}{\matS_b} 
\safemath{\dicta}{\matA} 	
\safemath{\dictb}{\matB} 	
\safemath{\hollowS}{H}
\safemath{\hollowA}{H_a}
\safemath{\hollowB}{H_b}
\safemath{\cross}{Z}
\safemath{\coh}{\mu_d}			
\safemath{\coha}{\mu_a}			
\safemath{\cohb}{\mu_b}			
\safemath{\mubs}{\nu}	
\safemath{\cohm}{\mu_m} 
\safemath{\dictset}{\setD}	
\safemath{\dictsetp}{\dictset(\coh,\coha,\cohb)}	
\safemath{\dictsetgen}{\dictset_\text{gen}}
\safemath{\dictsetgenp}{\dictsetgen(\coh)}
\safemath{\dictsetonb}{\dictset_\text{onb}}
\safemath{\dictsetonbp}{\dictsetonb(\coh)}

\safemath{\leftside}{U}
\safemath{\rightsideA}{R_a}
\safemath{\rightsideB}{R_b}

\safemath{\indexS}{\setI_S} 

\safemath{\na}{n_a}			
\safemath{\nb}{n_b}			
\safemath{\coeffa}{p_i}	
\safemath{\coeffb}{q_j}	
\safemath{\seta}{\setP}		
\safemath{\setb}{\setQ}     
\safemath{\setw}{\setW}	
\safemath{\setz}{\setZ}	
\safemath{\cola}{\veca}		
\safemath{\colb}{\vecb}		
\safemath{\cold}{\vecd}		
\safemath{\inputvec}{\vecx} 	
\safemath{\error}{\vece}	
\safemath{\noiseout}{\vecz} 	
\safemath{\inputvecel}{x}
\safemath{\inputveca}{\vecx_a}
\safemath{\inputvecb}{\vecx_b}
\safemath{\outputvec}{\vecy}	
\safemath{\lambdamin}{\lambda_{\mathrm{min}}}

\safemath{\elltwo}{\ell_2}
\safemath{\ellone}{\ell_1}
\safemath{\ellzero}{\ell_0}
\safemath{\ellinf}{\ell_\infty}
\safemath{\ellinftilde}{\ell_{\widetilde\infty}}
\safemath{\licard}{Z(\coh,\coha,\cohb)}
\safemath{\xsol}{\hat{x}}
\safemath{\xbord}{x_b}		
\safemath{\xstat}{x_s}		
\safemath{\xstatLone}{\tilde{x}_s}
\safemath{\order}{\mathcal{O}} 
\safemath{\scales}{\Theta} 
\safemath{\ones}{\mathbf{1}} 
\safemath{\zeroes}{\mathbf{0}} 
\safemath{\thlone}{\kappa(\coh,\cohb)} 
\safemath{\constoneA}{\delta} 
\safemath{\constoneB}{\epsilon} 
\safemath{\nlarge}{L}				   
\safemath{\sumlarge}{S_\nlarge}
\safemath{\maxlarger}{P_\nlarge}	   
\safemath{\Pzero}{\textrm{P0}}	
\safemath{\Pone}{\textrm{P1}}
\safemath{\vecfir}{\vecw}			 
\safemath{\vecsec}{\vecz}
\safemath{\elvecfir}{w}              
\safemath{\elvecsec}{z}				 
\safemath{\nlargefir}{n}
\safemath{\normout}{\gamma}
\safemath{\auxfun}{h}
\safemath{\supp}{\textrm{supp}}

\safemath{\indexa}{\ell}
\safemath{\indexb}{r}
\safemath{\indexc}{i}
\safemath{\indexd}{j}

\safemath{\project}{P}